\documentclass[showpacs,preprintnumbers,amsmath,amssymb,nofootinbib]{revtex4}
\usepackage{color}
\usepackage{epsf}
\usepackage{amssymb}
\usepackage{amsfonts}
\usepackage{latexsym}
\usepackage{mathrsfs}
\usepackage{graphicx,epsf}
\def\be{\begin{equation}}
\def\ee{\end{equation}}
\def\bea{\begin{eqnarray}}
\def\eea{\end{eqnarray}}

\def\R{{\cal{R}}}
\def\P{{\cal{P}}}
\def\S{{\cal{S}}}
\def\C{{\cal{C}}}
\def\O{\mathcal{O}}

\def\D{{\Delta}}
\def\ds{\delta s}

\def\esigsig{\eta_{\sigma\sigma}}
\def\dotesigsig{\dot{\eta}_{\sigma\sigma}}
\def\esigs{\eta_{\sigma s}}
\def\dotesigs{\dot{\eta}_{\sigma s}}
\def\ess{\eta_{ss}}
\def\dotess{\dot{\eta}_{ss}}
\def\e{\epsilon}
\def\eh{\epsilon^H}
\def\doteh{\dot{\epsilon}^H}
\def\ddoteh{\ddot{\epsilon}^H}
\def\xisigsigsig{\xi_{\sigma\sigma\sigma}^2}
\def\xisigsigs{\xi_{\sigma\sigma s}^2}
\def\xisigss{\xi_{\sigma s s}^2}
\def\xisss{\xi_{sss}^2}
\def\dk{d\log k}
\def\Tss{T_{\S\S}}
\def\Trs{T_{\R\S}}
\def\s{\sin\Delta\,}
\def\c{\cos\Delta\,}
\def\ss{\sin^2\!\!\Delta\,}
\def\cs{\cos^2\!\!\Delta\,}
\def\sc{\sin^3\!\!\Delta\,}

\def\cf{\cos^4\!\!\Delta\,}
\def\t{\tan\Delta\,}
\def\ts{\tan^2\!\!\Delta\,}
\def\({\left(}
\def\){\right)}
\def\nn{\nonumber}
\def\m{8\pi G}
\def\dphi{\delta\phi}
\def\0{^{(0)}}
\def\1{^{(1)}}
\def\2{^{(2)}}
\def\dTo{\frac{d T_{\R\S}^{(1)}}{\dk}}
\def\dTt{\frac{d T_{\R\S}\2}{\dk}}
\def\dlogTo{\frac{d \log T_{\R\S}^{(1)}}{\dk}}
\def\dlogTt{\frac{d \log T_{\R\S}\2}{\dk}}
\def\dlogTsso{\frac{d \log T_{\S\S}\1}{\dk}}
\def\dlogTsst{\frac{d \log T_{\S\S}\2}{\dk}}
\def\dao{\frac{d a_1}{d\log k}}
\def\dat{\frac{d a_2}{d\log k}}
\def\dathree{\frac{d a_3}{d\log k}}
\def\Pzero{\mathcal{P}^{(0)}_*}
\def\none{n_{(0)}^{(1)}}
\def\ntwo{n_{(0)}^{(2)}}

\def\nado{n_{\mathrm{ad}1}}
\def\nadt{n_{\mathrm{ad}2}}
\def\ncor{n_{\mathrm{cor}}}
\def\niso{n_{\mathrm{iso}}}

\begin{document}
\title{Curvature and isocurvature perturbations from two-field
  inflation in a slow-roll expansion}

\author{Christian T.~Byrnes and David Wands}

\affiliation{Institute of Cosmology and Gravitation, University of Portsmouth, Mercantile
House, Portsmouth~PO1~2EG, United Kingdom}
\date{\today}

\begin{abstract}
We calculate the power spectra of primordial curvature and isocurvature perturbations
from a general two field inflation model at next-to-leading order correction in a
slow-roll expansion. In particular we calculate the spectral indices to second order in
slow-roll parameters. We show that the cross-correlation of the curvature and
isocurvature perturbations at the time of Hubble-exit during inflation is non-zero at
first-order in slow-roll parameters. We apply our results to different classes of
inflation, including inflaton and curvaton scenarios. The spectrum of primordial
gravitational waves, curvature and isocurvature perturbations obey generalised
consistency relations in two-field inflation models. We give the first two consistency
relations in an infinite hierarchy.
\end{abstract}

\pacs{98.80.Cq \hfill astro-ph/0605679}

\maketitle

\section{Introduction}

During inflation the vacuum fluctuations of light scalar fields grow into super-Hubble
density perturbations which are believed to be the origin of the structure seen in the
universe today. Single-field slow roll inflation predicts a nearly scale invariant
spectrum of gaussian curvature perturbations. However from a particle physics point of
view it is natural to expect there to be more then one scalar field rolling during
inflation. During multiple field inflation a spectrum of isocurvature as well as
curvature perturbations can be generated and the two may be correlated \cite{langlois}.

In this paper we consider the case of two field inflation with an arbitrary potential and
arbitrary background trajectory. Following Ref.~\cite{AGWS,wands} we introduce arbitrary
transfer functions to parameterise the generation of primordial curvature and
isocurvature perturbations from isocurvature perturbations during inflation.  We present
the power spectra of the isocurvature and curvature perturbations to first-order in slow
roll as well as the cross-correlation, presenting the results in terms of the slow-roll
parameters and the (in principle) observable correlation angle. We are then able to
calculate the tilts of all three power spectra to second-order in slow-roll parameters
(see also Ref.~\cite{vanTent}). This generalises the result of Stewart and Lyth
\cite{stewartlyth} for adiabatic perturbations from single-field inflation. For an
alternative approach to calculating the primordial curvature perturbation in multi-field
models (but not the isocurvature perturbation), based on the $\delta
N$-formalism~\cite{sasakistewart}, see Ref.~\cite{Lee:2005bb,gong}.

In section \ref{initialpowerspectra} we calculate the scalar field perturbations during
inflation, first performing a global rotation in field space to a basis where the
perturbations are uncorrelated at Hubble exit. We then rotate to the local
(instantaneous) basis of adiabatic and isocurvature perturbations
\cite{gordon,nibbelink,vanTent} and show that they are correlated at first-order in slow
roll at Hubble exit. The only exception is the case of a straight background trajectory
in field space, e.g.\,for a symmetric potential~\cite{byrnes}, when the local rotation is
in fact a constant (global) rotation which coincides with the uncorrelated basis and
hence the curvature and isocurvature perturbations are uncorrelated at Hubble exit.

In section \ref{observables} we change variables to the dimensionless curvature and
isocurvature perturbations which are then related to the primordial curvature and
isocurvature perturbations after inflation, and introduce the dimensionless correlation
angle. The curvature and isocurvature perturbations are evolved on superhorizon scales by
introducing two arbitrary transfer functions, which parameterise our ignorance of the
detailed physics after Hubble exit and through the end of inflation and reheating. We
present results for the power spectra and tilts including the next to leading order terms
and also the running at leading order, which is second-order in slow roll. We also
briefly present the power spectrum, tilt and running for tensor perturbations, which are
unchanged from the single-field case \cite{Lidsey:1995np}. In section
\ref{modeldependentrelations} we simplify the results to the interesting special cases of
a straight background trajectory, which includes assisted inflation with exponential
potentials \cite{liddle,malik,copeland}, the curvaton scenario \cite{curvaton}, and the
inflaton scenario allowing for an additional uncorrelated isocurvature field present
during inflation. We conclude in section \ref{conclusions}.

Appendices contain definitions of all the slow-roll parameters used and the relations
between these parameters when defined in terms of the Hubble parameter or the potential,
their derivatives and further details of the calculation of the tilts. Throughout this
paper we use the notation that results accurate at first order in slow roll are denoted
by $\simeq$, while equality at second order in slow roll is denoted by $\cong$.

\section{Initial Power spectra}\label{initialpowerspectra}

\subsection{Background equations of motion}

We consider two scalar fields with Lagrange density
 \be \mathcal{L} =
 -\frac12\sum_{I=1}^{2} g^{\mu\nu} \phi_{I,\mu}\phi_{I,\nu}-V(\phi_1,\phi_2).
 \ee
We thus allow for an arbitrary interaction potential,
 $V(\phi_1,\phi_2)$, but for simplicity consider only canonical
 kinetic terms. It should be possible to generalise our
 results
 to non-minimal kinetic terms, see for example \cite{nibbelink,vanTent}.

The background equations of motion are
 \bea &&\ddot{\phi}_I+3H\dot{\phi}_I+V_I=0, \\ &&
H^2=\frac{\m}{3}\left[V+\frac12\(\dot{\phi}_1^2+\dot{\phi}_2^2\)\right] \,,
 \eea
where $V_I\equiv \partial V/\partial\phi_I$.
To solve these equations we use the slow roll approximation to rewrite them as
\bea \dot{\phi}_I \simeq -\frac{V_I}{3H}\(1+\frac13 \delta_I^H\)\,, \qquad
 H^2 \simeq \frac{\m}{3}V\(1+\frac13\eh\)\,, \eea
where $\delta_I^H$ and $\eh$ are slow-roll parameters as defined in appendix
\ref{slowrolldefinition}, which we assume to be small. We make the standard slow roll
assumption that the time derivative of the slow-roll parameters are higher order in slow
roll.

\subsection{The perturbation equations}

The equations of motion for the perturbed fields in the spatially flat gauge are
\cite{Taruya:1997iv}
\be \ddot{\dphi}_I+3H\dot{\dphi}_I+\frac{k^2}{a^2}\dphi_I+
  \sum_J
   \left[V_{IJ}-\frac{\m}{a^3}\frac{d}{dt}\(\frac{a^3}{H}\dot{\phi}_I\dot{\phi}_J\)\right]
   \dphi_J=0
 \,,
 \ee
where $k$ is the comoving wavenumber. To simplify the above equations we change variables
to $u_I=a\dphi_I$ and to conformal time $\tau$, defined by $d\tau=dt/a$. To lowest order
in slow roll,
$\tau\simeq -(1+\e)/(aH)$, the equations of motion simplifies to
\be \label{ueom} u_I^{''}+\(k^2-\frac{2}{\tau^2}\)u_I=\frac{3}{\tau^2}\sum_J M_{IJ}u_J
\ee
where the prime denotes a derivative with respect to conformal time and the interaction
matrix is
\be \label{MIJ} M_{IJ}\simeq\left(\begin{array}{cc} \eh+2\eh_{11}-\eta_{11} &
2\eh_{12}-\eta_{12}
\\ 2\eh_{12}-\eta_{12}&\eh+2\eh_{22}-\eta_{22}
\end{array} \right). \ee
Note that the superscript $H$ denotes a slow-roll parameter defined in terms of the
Hubble parameter, while all other slow-roll parameters are defined in terms of the
potential (see appendix \ref{slowrolldefinition}).

The two differential equations for $u_I$ are coupled but they can be decoupled at
Hubble-exit by performing a rotation to diagonalise the interaction matrix (\ref{MIJ}) to
first order in slow roll. For previous discussions on how to solve the coupled equations
(\ref{ueom}) see, for example, \cite{bmr,nibbelink}. The rotation matrix is given by
\bea \label{U} U= \(\begin{array}{cc} \cos\Theta & -\sin\Theta \\
\sin\Theta & \cos\Theta \end{array}\) \eea
where the rotation angle $\Theta$ is defined by
\bea \label{tanTheta} \tan 2\Theta
= 2\left[ \frac{2\eh_{12}-\eta_{12}}{2(\eh_{11}-\eh_{22})-(\eta_{11}-\eta_{22})}
\right]_*\,. \eea
The subscript $*$ refers to a quantity evaluated at horizon crossing, $k=a_* H_*$. For a
given wavenumber $k$, $\Theta$ is a constant, but there is a $k$ dependence since the
slow roll parameters are evaluated at horizon crossing and each wave mode has a different
horizon crossing time.  Note that since the rotation angle $\Theta$ is defined in terms
of a ratio of slow-roll parameters it is in general not small. If all of the slow-roll
parameters become arbitrarily small Eq.~(\ref{tanTheta}) becomes ambiguous, because the
equations of motion (\ref{ueom}) are then decoupled to first order in any basis and any
rotation will work. The rotation matrix (\ref{U}) satisfies
\be U^\dag MU=\mathrm{diag}(\lambda_1,\lambda_2), \ee
where
\be \label{lambdaIJ} \lambda_{1,2}\simeq \frac12 \left[ 4\eh-(\eta_{11}+\eta_{22})
\pm\sqrt{[2(\eh_{11}-\eh_{22})-(\eta_{11}-\eta_{22})]^2+4(2\eh_{12}-\eta_{12})^2}
\right]_*. \ee
Denoting the decoupled variables by $v_I$, we have
\be u_I= \sum_J U_{IJ} v_J \,. \ee
Therefore left-multiplying (\ref{ueom}) by $U^\dag$ and rewriting in the $v_I$ basis we
have \bea &&
v_I^{''}+\left(k^2-\frac{1}{\tau^2}(\mu_I^2-\frac14)\right)v_I
 \simeq 0 \,,
\eea where \be \mu_{I} = \frac32+\lambda_I \,. \ee This is the same form for the equation
of motion as in the single field case \cite{stewartlyth} except that the $\mu_I$ depend
on slow-roll parameters relating to both the original fields $\phi_1$ and $\phi_2$.

Working at first order in slow roll, so that we can treat the slow-roll parameters and
hence $\mu_I$ as constant, the solution is
\be v_I = \frac{\sqrt{\pi}}{2}e^{i(\mu_I+1/2)\pi/2}(-\tau)^{1/2}
H^{(1)}_{\mu_I}(-k\tau)e_I(k) \ee where $H_{\mu_I}^{(1)}(x)$ is a Hankel function of the
first kind of order $\mu_I$ and the $e_I(k)$ are independent unit Gaussian random
variables \cite{gordon} satisfying \be \langle e_I({\bf k}) \rangle = 0 \,, \qquad
\langle e_I({\bf k})e_J^*({\bf k'})\rangle
 = \delta_{IJ}\delta^3(\bf{k}-\bf{k'}) \,.
\ee
The constant in front of the Hankel function solution for $v_I$ comes from imposing the
early-time boundary condition
\be v_I \longrightarrow\frac{1}{\sqrt{2k}}e^{-ik\tau}e_I(k) \qquad \mathrm{as} \qquad
-k\tau\longrightarrow \infty \,. \ee
The corresponding late-time behavior ($k\tau\to0$) is
\bea v_I\longrightarrow && e^{i(\mu_I-1/2)\pi/2}2^{\mu_I-3/2}
\frac{\Gamma(\mu_I)}{\Gamma(3/2)}\frac{1}{\sqrt{2k}}(-k\tau)^{-1-\lambda_I}e_I(k)
 \nonumber\\
\label{vlatetime}
 \simeq && ie^{i\lambda_I\pi/2}
(1+C\lambda_I)\frac{1}{\sqrt{2k}}(-k\tau)^{-1-\lambda_I}e_I(k) \eea
where $C=2-\log2-\gamma\approx0.7296$ and $\gamma\approx0.5772$ is the Euler-Mascheroni
constant. Note however that although the change in first-order slow-roll parameters and
hence $\mu_I$ in one Hubble time is second order in slow roll, see appendix
\ref{derivatives}, the time variation of the slow-roll parameters cannot be neglected
over many Hubble times. Thus we can only reliably use the late-time solution
(\ref{vlatetime}) for a few Hubble times after Hubble-exit ($k=a_* H_*$). We need to
perform another rotation in field space to accurately track the evolution of the field
perturbations over many expansion times on super-Hubble scales.

\subsection{The adiabatic and isocurvature perturbations}

To follow the coupled evolution of the perturbations on large scales ($k\ll aH$) it is
more useful to rotate the field basis into adiabatic and entropy perturbations, $\delta
\sigma$ and $\delta s$. Following Ref.\,\cite{gordon}, the local rotation is given by
\bea
\label{thetarotation} \(\begin{array}{c} \delta\sigma \\
\delta s
\end{array}\)=S^\dag \(\begin{array}{c} \delta\phi_1 \\ \delta \phi_2 \end{array}\)
\qquad
{\rm where}\qquad S=\(\begin{array}{cc} \cos\theta & -\sin\theta \\
\sin\theta & \cos\theta \end{array}\) \eea
and the local rotation angle $\theta$, which is now a function of time, is given by \bea
\label{tantheta} \tan\theta=\frac{\dot{\phi}_2}{\dot{\phi}_1} \,. \eea The adiabatic and
isocurvature perturbations
are then related to the decoupled perturbations $v_I$ by
\bea
\(\begin{array}{c}a\delta\sigma \\
a\delta s\end{array}\)
=\(\begin{array}{cc} \cos(\Theta-\theta)&-\sin(\Theta-\theta) \\
\sin(\Theta-\theta)& \cos(\Theta-\theta)\end{array}\)\(\begin{array}{c} v_1 \\ v_2
\end{array} \).
\eea
The eigenvalues of $U$, given in Eq.~(\ref{lambdaIJ}) can be written more compactly in
terms of the $\sigma$ and $s$ slow-roll parameters (see appendix
\ref{slowrolldefinition})
\bea
\lambda_{1,2}\simeq\frac12\left[4\e-(\esigsig+\ess)\pm\sqrt{\varpi^2+4\esigs^2}\right]_*
\,, \eea
where
\bea \varpi=2\e-(\esigsig-\ess)\,, \eea
and the global rotation angle $\Theta$ defined in Eq.~(\ref{tanTheta}) simplifies to
\bea\label{tanThetasigs} \tan2\Theta =
\left[\frac{\varpi\sin2\theta-2\esigs\cos2\theta}{\varpi\cos2\theta+2\esigs\sin2\theta}\right]_*
\,. \eea From the above equation it is easy to see that the two rotations, by $\Theta$
and $\theta$, are the same if and only if $\esigs=0$, i.e.~in the special case of a
straight background trajectory since \cite{wands} $\dot{\theta}\simeq-H\esigs$.

Since $v_1$ and $v_2$ are uncorrelated we have
\bea a^2\langle|\delta\sigma|^2\rangle
& = & \frac12 \(\langle v_1^2\rangle+ \langle v_2^2\rangle\)+ \frac12\cos2(\Theta-\theta)
\( \langle v_1^2\rangle-\langle v_2^2\rangle \)\,,  \\
a^2\langle\delta\sigma \delta s^*\rangle
& = & \frac12 \sin2(\Theta-\theta) (\langle v_1^2\rangle-\langle v_2^2\rangle)\,, \\
a^2\langle |\delta s|^2\rangle
& = & \frac12 \( \langle v_1^2\rangle+\langle v_2^2\rangle \) -\frac12
\cos2(\Theta-\theta) (\langle v_1^2\rangle-\langle v_2^2\rangle)\,, \eea
and from (\ref{tantheta}) and (\ref{tanThetasigs}) the trigonometric terms can be written
in terms of the slow-roll parameters
\bea \cos2(\Theta-\theta) =
\frac{\varpi}{\sqrt{\varpi^2+4\esigs^2}}+\mathcal{O}(\epsilon,\esigsig,\esigs,\ess),
\qquad \sin2(\Theta-\theta) =
\frac{-2\esigs}{\sqrt{\varpi^2+4\esigs^2}}+\mathcal{O}(\epsilon,\esigsig,\esigs,\ess) \,.
\eea
A key observation is that $\delta\sigma$ and $\delta s$ are correlated at Hubble exit at
first order in slow roll if the background trajectory is curved. Using the above
relationships and Eq.~(\ref{vlatetime}) we can calculate the power spectra and
cross-correlation at Hubble exit
\bea \label{Psig}  \P_{\sigma *}(k) & \simeq &
\left(\frac{H_*}{2\pi}\right)^2\(1+(-2+6C)\e-2C\esigsig\)\,,
\\
\label{Csigs} \C_{\sigma s*}(k)
& \simeq & -2C\esigs\left(\frac{H_*}{2\pi}\right)^2\,, \\
\label{Ps} \P_{s*}(k)&\simeq & \left(\frac{H_*}{2\pi}\right)^2 \(1+(-2+2C)\e-2C\ess\)\,,
\eea
where here, and in the following, the slow-roll parameters are to be evaluated at Hubble
exit, and we define the autocorrelation $\P_{x}\equiv\C_{xx}$ and the cross-correlation
\be \label{powerspectrumdefinition}
 \C_{xy} \delta^3(\mathbf{k}-\mathbf{k'}) \equiv \frac{4\pi k^3}{(2\pi)^3}
 \langle x(\mathbf{k}) y^*(\mathbf{k'}) \rangle \,.
\ee

\section{Observables and Final Power spectra}\label{observables}

The comoving curvature perturbation during inflation is given in terms of the adiabatic
field perturbations in the spatially flat gauge, by \cite{gordon}
\be \R=\frac{H}{\dot{\sigma}}\delta\sigma. \ee
Similarly a dimensionless isocurvature perturbation during inflation can be defined as
\cite{AGWS}
\be
 \label{defSstar}
\S=\frac{H}{\dot{\sigma}}\ds \,. \ee
A convenient dimensionless measure of the correlation angle $\D$ is
\be\label{correlationangle} \cos\D\equiv\frac{\C_{\R\S}}{\P_{\R}^{1/2}\P_{\S}^{1/2}}.
\ee

It is straightforward to convert the power spectra and cross-correlation at Hubble exit,
given in Eqs.~(\ref{Psig}--\ref{Ps}), to the dimensionless variables \bea \label{PR}
\P_{\R*} & \simeq &
\left(\frac{H^2_*}{\dot{\sigma}_*2\pi}\right)^2(1+(-2+6C)\e-2C\esigsig)=\Pzero(1+a_1),
\\ \label{CRS} \C_{\R\S *} & \simeq & -2C\esigs\left(\frac{H^2_*}{\dot{\sigma}_*2\pi}\right)^2
= \Pzero a_2,
\\ \label{PS} \P_{\S*} & \simeq & \left(\frac{H^2_*}{\dot{\sigma}_*2\pi}\right)^2(1+(-2+2C)\e-2C\ess)
= \Pzero(1+a_3), \\
\cos\D_* & \simeq & -2C\esigs\,, \eea
where we have defined
\bea \label{Pzero} \Pzero \equiv \left(\frac{H^2_*}{\dot{\sigma}_*2\pi}\right)^2=
\left(\frac{H_*}{2\pi}\right)^2\frac{16\pi G}{\eh}\,, \eea
and the first-order slow-roll corrections are given by
\bea \label{aI}  a_1=-2\epsilon+2C(3\epsilon-\esigsig)\,, \qquad a_2=-2C\esigs\,, \qquad
a_3=-2\epsilon+2C(\epsilon-\ess)\,. \eea

\subsection{Super-Hubble evolution}

In order to calculate the primordial curvature and isocurvature perturbations some time
after inflation has ended, which can be constrained by observations today we need to
model the evolution of $\R$ and $\S$ on large scales by introducing the transfer
functions
\bea
 \label{Tmatrix}
\(
\begin{array}{c} \R \\
\S \end{array}\)=\( \begin{array}{cc}1 & \Trs \\ 0 & \Tss \end{array} \)
\(\begin{array}{c} \R \\ \S \end{array}\)_*. \eea
$T_{\R\R}=1$ because in the absence of isocurvature modes the adiabatic perturbation is
conserved and $T_{\S\R}=0$ because the adiabatic perturbation cannot act as a source to
the isocurvature perturbation.

On the other hand, $T_{\R\S}$ and $T_{\S\S}$, which parameterise the effect of entropy
perturbations during inflation upon the primordial curvature and isocurvature
perturbations, depend upon the full evolution on super-Hubble scales, both during
inflation, and afterwards. Quite generally, we have
\bea &&\dot{\R}=\alpha(t) H \S, \qquad \dot{\S}=\beta(t) H \S\,, \eea
and thus we can write \cite{wands}
\bea \Tss(t_*,t)=\exp\(\int_{t_*}^t \beta(t')H(t')d t' \), \qquad
\Trs=\int_{t_*}^t\alpha(t')\Tss(t_*,t')H(t')dt'. \eea
The transfer functions contain an implicit scale dependence, through the dependence of
the Hubble-exit time, $t_*$, which varies with scale, $k=(aH)_*$. The time dependence,
calculated using the Leibniz integral rule, is
\be \label{Tderiv} \frac1H_*\frac{\partial \Tss}{\partial t_*}=-\beta_*\Tss, \qquad
\frac1H_*\frac{\partial \Trs}{\partial t_*}=-\alpha_*-\beta_*\Trs. \ee To calculate the
spectral indices of the primordial power spectra to second order in slow roll it is thus
necessary to also calculate $\alpha_*$ and $\beta_*$ to second order.

On large scales the time derivative of the curvature perturbation is~\cite{bassett}
\be \label{Rdot} \dot{\R}=2\dot{\theta}\S \,. \ee
Differentiating Eq.~(\ref{tantheta}) for $\tan\theta$ gives
\be \dot{\theta}=H\sin\theta\cos\theta(\delta_1^H-\delta_2^H) \,, \ee
where $\delta_I^H$ are slow-roll parameters defined in appendix \ref{relating}. Then
using appendix \ref{relating} to write $\delta_I^H$ in terms of the potential slow-roll
parameters we find
\be \label{alpha} \alpha_* \cong
\(-2+\frac83\e-\frac23\esigsig-\frac23\ess\)\esigs-\frac23\xisigsigs \, . \ee
As can be easily seen from (\ref{Rdot}), $\alpha_*=0$ in the case of a straight
background trajectory in field space, and the curvature perturbation in that case is
constant during inflation on super-Hubble scales.

To calculate $\beta_*$ we start with the evolution equation of $\ds$ \cite{gordon}
\bea \label{seom} \ddot{\ds}+3H\dot{\ds}+\(\frac{k^2}{a^2}+V_{ss}+3\dot{\theta}^2\)\ds =
\frac{\dot{\theta}}{\dot{\sigma}}\frac{k^2}{2\pi G a^2}\Psi, \eea
where the Bardeen potential $\Psi\simeq 4\pi G(\dot\sigma/H)\delta\sigma$ during
slow-roll on large scales \cite{gordon}.
Therefore we can drop the term on the right hand side of (\ref{seom}) as well as the
$(k^2/a^2)\ds$ term on large scales.  Rewriting the equation in terms of $\S$ we then
find
\be\label{Sfulleom} \ddot{\S}+\(3H+\frac{\doteh}{\eh}\)\dot{\S}+\(
V_{ss}+3\dot{\theta}^2+\frac32H\frac{\doteh}{\eh}+\frac12\frac{\ddoteh}{\eh}-\frac14
\(\frac{\doteh}{\eh}\)^2 \)\S=0, \ee
so, to lowest order in slow roll, we have
\be \dot{\S} \simeq H(-2\epsilon+\esigsig-\ess)\S \,. \ee
We are able to calculate the next order corrections by taking the derivative of this and
substituting it back into (\ref{Sfulleom}). After some calculation, we find
\be \label{beta} \beta_* \cong \frac13 \left( -6\epsilon +3\esigsig- 3\ess
+12\epsilon^2+\esigsig^2-\ess^2 -10\epsilon\esigsig+
2\epsilon\ess+\xisigsigsig-\xisigss\right). \ee

\subsection{Final Power Spectra and spectral indices}

The primordial curvature perturbation, during the radiation-dominated era some time after
inflation has ended, is given on large scales by
\be \R=\psi+\frac{H\delta\rho}{\dot{\rho}} \,. \ee
It is this curvature perturbation that, for example, produces large-scale anisotropies in
the cosmic microwave background. The conventional definition of the primordial
isocurvature matter perturbation is given relative to the radiation density by
\be \S = H \left( \frac{\delta\rho_m}{\dot{\rho}_m} -
 \frac{\delta\rho_\gamma}{\dot{\rho}_\gamma} \right)
\ee
From the definition of the power spectra (\ref{powerspectrumdefinition}) and the transfer
functions (\ref{Tmatrix}) it follows that
\bea \label{PRfinalsimple} \P_{\R} & \simeq & \P_{\R *} +2T_{\R\S}\C_{\R\S *} +T_{\R\S}^2
\P_{\S *}  \,,
\\ \label{CRSfinalsimple} \C_{\R\S} & \simeq & T_{\S\S}\C_{\R\S *}
 +T_{\S\S}T_{\R\S} \P_{\S *} \,, \\
\label{PSfinalsimple}  \P_{\S} & \simeq &  T_{\S\S}^2 \P_{\S *} \,, \eea

and substituting in the power spectra at Hubble exit (\ref{PR}--\ref{PS}) we see that the
primordial power spectra are
\bea    \label{PRfinal} \P_{\R} & \simeq &
\Pzero(1+T_{\R\S}^2+a_1+2T_{\R\S}a_2+T_{\R\S}^2a_3)\,,
\\ \label{CRSfinal} \C_{\R\S} & \simeq & \Pzero T_{\S\S}(T_{\R\S}+a_2+T_{\R\S}a_3)\,,
\\
\label{PSfinal}  \P_{\S} & \simeq & \Pzero T_{\S\S}^2(1+a_3)\,, \eea where $a_1,a_2$ and
$a_3$ are the corrections from $\Pzero$ for the initial power spectra as defined in
Eqs.~(\ref{Pzero}) and (\ref{aI}).

Note that the next-to-leading order corrections to the power spectra
(\ref{PRfinal}--\ref{PSfinal}) can be related to the tilts by the relation given in
Ref.~\cite{vanTent}
\bea \P_X\simeq \P_X\0\left(1+n_T-Cn_X\right), \eea
where the subscript $X$ means that that the relation is true for the adiabatic,
isocurvature and tensor power spectra and the cross correlation, and $\P_X\0$ is the
power spectra at lowest order. The tilt of the gravitational wave power spectrum, $n_T$,
is defined in section \ref{gravitationalwaves}

The scale dependence depends on both the initial power spectra and the transfer
functions, we can replace the dependence on the transfer functions with the observable
correlation angle $\c\!$, which at lowest order satisfies
\bea \c\!\0=\frac{T_{\R\S}}{\sqrt{1+T_{\R\S}^2}}. \eea
For details of the calculation and definitions of $\c\!,\s\!,\t\!$ see appendix
\ref{tiltcalculation}.

The final scalar tilts, defined as $n_X=d\ln \P_X/d\ln k$, up to second order in
slow-roll parameters are\footnote{In this notation a scale-invariant
(Harrison-Zel'dovich) spectrum corresponds to $n_\R=0$.}
\bea  \label{nR} n_\R & \cong & -(6-4\cs)\e+2\ss\esigsig+4\s\c\esigs+2\cs\ess \\
\nn &&+
 \(-\frac{10}{3}-4\cs+C\(24-16\cs\)\)\e^2 +\frac23\ss\esigsig^2+\(\frac23\cs+4C(1-2\cs)\)\esigs^2
\\ \nn &&
+\frac23\cs\ess^2 + \(-2+\frac{14}{3}\cs+C\(-16+12\cs\)\)\e\esigsig-\frac43\s\c(1+6C)\e\esigs \\
\nn && -\frac23\cs(-1+6C)\e\ess+\frac43\s\c(1-3C)\esigsig\esigs + \frac43\s\c(1+3C)\esigs\ess \\
\nn && +\frac23\ss(1+3C)\xisigsigsig +
\frac43\s\c\(1+3C\)\xisigsigs+\frac23\cs\(1+3C\)\xisigss\,, \\
\label{nC} n_\C & \cong & -2\e+2\t \esigs +2\ess\\
\nn && +\(-\frac{22}{3}+8C\)\e^2+\frac23\left(1-6C\right)\esigs^2+\frac23\ess^2
+\frac83\left(1-\frac32C\right)\e\esigsig-\frac23(1+6C)\t\e\esigs +\frac23(1-6C)\e\ess
\\ \nn &&
+\frac23(1-3C)\t\esigsig\esigs+\frac23(1+3C)\t\esigs\ess+
\frac23(1+3C)\t\xisigsigs+\frac23(1+3C)\xisigss\,,
\\ \label{nS}
n_\S & \cong & -2\e+2\ess \\
\nn && +\(-\frac{22}{3}+8C \)\e^2 +
\frac23\left(1-6C\right)\esigs^2+\frac23\ess^2+\frac83\left(1-\frac32C\right)\e\esigsig +
\frac23\left(1-6C\right)\e\ess +\frac23\left(1+3C\right)\xisigss\,. \eea

The running of the spectral index is defined as $\alpha_X \equiv dn_X/d\ln k$. Assuming
that the power law approximation for $\P_\R, \C_{\R\S}$ and $\P_\S$ are valid, the
running will be second order in the slow roll approximation because the time derivatives
(or equivelantly $\ln k$ derivatives) of the first-order slow-roll parameters are second
order. To leading order in slow roll (see also \cite{vanTent}),
\bea  \label{alphaR} \alpha_\R&\cong&8(-3+4\cs-2\cf)\e^2+4\ss\cs\esigsig^2+4(1-4\ss\cs)\esigs^2+4\ss\cs\ess^2 \\
\nn && +4(4-7\cs+4\cf)\e\esigsig+32\sc\c\e\esigs+4\cs(5-4\cs)\e\ess \\
\nn && -8\s\c(1-2\cs)\esigsig\esigs -8\ss\cs\esigsig\ess +8\s\c(1-2\cs)\esigs\ess  \\ \nn
&& -2\ss\xisigsigsig -4\s\c\xisigsigs -2\cs\xisigss\,,
\\
\label{alphaC} \alpha_C&\cong&-8\e^2 +4(1-\ts)\esigs^2
+4\e\esigsig+4\e\ess+4\t\esigsig\esigs-4\t\esigs\ess-2\t\xisigsigs-2\xisigss\,,
\\ \label{alphaS} \alpha_\S&\cong&-8\e^2+4\esigs^2+4\e\esigsig+4\e\ess-2\xisigss\,. \eea
It would be a straightforward but long calculation to include the next to leading terms
also, since all that is required is to differentiate the tilts including the leading
corrections as given above.

The spectral indices of $\P_\R$ and $\P_\S$ are both slow roll suppressed while
generically $\alpha_{\R/\S}\sim\mathcal{O}(n_{\R/\S}^2)$ so the power spectra are both
weakly scale dependent and well approximated by power laws. However the cross-correlation
can be strongly scale-dependent if $\t$ is large, specifically if
\be \t \esigs\sim\O(1). \ee
In this case the running $\alpha_C\sim\O((\t\esigs)^2)$ is also large and so $\C_{\R\S}$
is not well approximated by a power law. The problem is that $T_{\R\S}$ does not have a
power law shape whenever $|\alpha_*| \gtrsim |\beta_* T_{\R\S}|$ as shown by
(\ref{Tderiv}) and then $\C_{\R\S}$ will not be close to power law either
(\ref{CRSfinalsimple}). We can see in this case $C_{\R\S}$ has an approximate
log-dependence on wavenumber from (\ref{Tderiv}) and (\ref{CRSfinal}). Fortunately this
problem only occurs when $\c\approx 0$, i.e.~the perturbations are nearly uncorrelated
and then $n_\C$ will be nearly unconstrained by observations. Because $T_{\R\S}$ has to
be small when it is not a power law (recalling that $\t\sim 1/T_{\R\S}$), $\P_\R$ remains
a power law in spite of its leading order dependence on $T_{\R\S}^2$
(\ref{PRfinalsimple}). We can see this more explicitly by parameterizing $\P_\R$ as the
sum of two power laws which we discuss next.

\subsection{Alternative parametrisation of the power spectra}

A different way to write the primordial power spectra is to split the adiabatic power
spectrum into a part generated by the inflationary adiabatic perturbations and a second
adiabatic perturbation generated from the inflationary entropy perturbation
\cite{valiviita}, (see also \cite{AGWS,WMAP}),
\bea \label{PRjussi} \P_\R & = &
A_r^2\(\frac{k}{k_0}\)^{\nado}+A_s^2\(\frac{k}{k_0}\)^{\nadt},
\\ \C_{\R\S} &=& A_sB\(\frac{k}{k_0}\)^{\ncor}, \\
\P_\S &=& B^2\(\frac{k}{k_0}\)^{\niso},  \eea
where $k_0$ is the pivot scale. The amplitude of the primordial curvature perturbation
spectra are given in terms of the power spectra at Hubble exit and the primordial
transfer functions (see Eq.~(\ref{PRfinalsimple})) by
\bea A_r^2=\left[ \P_{\R *} \right]_{k_0}, \qquad A_s^2=\left[ T_{\R\S}^2\P_{\S
*}+2T_{\R\S}\C_{\R\S *}\right]_{k_0}. \eea
$A_s^2$ and $A_r^2$ can also be simply written in terms of the correlation angle,
(\ref{correlationangle}),
\bea \label{correlationanglejussi} A_r^2=\left[ \P_{\R}\ss \right]_{k_0}, \qquad
A_s^2=\left[ \P_{\R}\cs \right]_{k_0}. \eea
Only 3 of the four tilts are independent because $\ncor=(\nadt+\niso)/2$. $\C_{\R\S}$ and
$\P_\S$ are the same as in the standard notation so $\ncor=n_\C$, $\niso=n_\S$ and
$\nadt=2n_\C-n_\S$. $\nado$ is the tilt of the adiabatic perturbations at the Hubble-exit
time of $k_0$. The four spectral indices are therefore
\bea\label{nado} \nado &\cong& -6\epsilon+2\esigsig
+\(-\frac{10}{3}+24C\)\epsilon^2+\frac23\esigsig^2 +\frac23(1+6C)\esigs^2
-(2+16C)\epsilon\esigsig+\(\frac23+2C\)\xisigsigsig, \\ \label{nadt} \nadt &\cong&
-2\epsilon+4\t\esigs+2\ess+\(-\frac{22}{3}+8C\)\epsilon^2
+\frac23(1-6C)\esigs^2+\frac23\ess^2 +\(\frac83-4C\)\epsilon\esigsig  \\  \nonumber &&
-\(\frac43+8C\)\t\epsilon\esigs +\(\frac23-4C\)\epsilon\ess +
\frac43(1-3C)\t\esigsig\esigs +\frac43(1+3C)\t\esigs\ess  \\ \nn &&
+\frac43(1+3C)\t\xisigsigs+\frac23(1+3C)\xisigss\,, \eea
while $\ncor=n_\C$ and $\niso=n_\S$ are written explicitly in (\ref{nC}) and (\ref{nS})
respectively.

In the case when $\t\esigs$ is large $\nadt$ is large and so is
$\alpha_{\mathrm{nad}_2}=2\alpha_\C-\alpha_\S$. Hence the second term of the primordial
adiabatic power spectrum (\ref{PRjussi}) is not well parameterised by power law. However
$\P_\R$ can still be well approximated by a single power law because $A_s^2\ll A_r^2$
from (\ref{correlationanglejussi}) and the requirement that $|\t|\gg1$.

\subsection{Gravitational Waves}\label{gravitationalwaves}

Scalar and tensor perturbations are decoupled, so the gravitational wave power spectrum
is the same as in the single-field result, and the amplitude of gravitational waves
remains frozen-in on large scales after Hubble exit during inflation, \cite{stewartlyth},
\be
 \P_T=\P_{T*}\simeq
64\pi G\(\frac{H_*}{2\pi}\)^2\(1+2(-1+C)\e\) \,. \ee
The tilt and running at second order in slow roll can easily be calculated from this
power spectrum to give
\bea n_T&\cong & -2\e\left[1+\(\frac{11}{3}-4C\)\e+\(-\frac43+2C\)\esigsig\right]\,,
\\ \alpha_T&\cong &-8\e^2+4\e\esigsig\,. \eea
The tensor-scalar ratio\footnote{Note that there are various definitions for the
tensor-scalar ratio, we use the definition which satisfies $r\simeq 16\e$ to first order
at Hubble exit as used in \cite{WMAP}, for example.} at Hubble exit is the same as in the
single-field case
\be \label{rstar} r_* \equiv \frac{\P_{\R*}}{\P_{T*}}\cong
16\e\left[1-\(\frac43+4C\)\e+\(\frac23+2C\)\esigsig\right] \,, \ee
However some time after Hubble exit the scalar curvature perturbation may have evolved
due to the effect of non-adiabatic perturbations. From the definition
(\ref{correlationangle}), the final power spectra (\ref{PRfinal}--\ref{PSfinal}) and the
curvature power spectrum at Hubble exit (\ref{PR}) it follows that \bea \P_{\R*}\cong
\P_\R\ss\,, \eea and thus in the radiation-dominated era we have
\bea \label{rfinal} r \cong
 16\e\ss\left[1-\(\frac43+4C\)\e+\(\frac23+2C\)\esigsig\right].
\eea Note that the only difference from the single field tensor-scalar relation is the
addition of the $\ss$ factor, this explains why the observational upper bound on $r$ does
not provide a direct upper bound on $\e$ in the multiple field case.

\section{model dependent relations}\label{modeldependentrelations}

\subsection{Straight background trajectory}

In the case of a straight background trajectory in field space (during the time of Hubble
exit for observable modes) the calculation simplifies considerably since then
$\dot{\theta}=0$. From (\ref{alpha}) this requires that the slow-roll parameters $\esigs=
\xisigsigs = 0$.
In the case of a symmetric potential \cite{byrnes} the background trajectory will be
straight since any orthogonal velocities will decay quickly. A straight background
trajectory is also the attractor solution of some assisted inflation models
\cite{liddle,malik,copeland}. The inflaton and curvaton scenarios are two classes of
inflation models with a straight background trajectory which we discuss in the following
two subsections.

There is a consistency relation independent of the gravitational wave background which
holds whenever the background trajectory is straight and which is valid at all orders in
slow roll. From (\ref{seom}) the adiabatic and entropy perturbations decouple exactly
when $\dot{\theta}=0$. Therefore the rotations performed before and after Hubble exit are
the same, $\Theta=\theta$, as was already shown to lowest order in slow roll from
(\ref{tanThetasigs}) and the adiabatic and isocurvature perturbations are uncorrelated at
Hubble exit. Equations (\ref{CRSfinalsimple}) and (\ref{PSfinalsimple}) with
$\C_{\R\S*}=0$ imply
\bea \label{straightcr} T_{\S\S}\C_{\R\S}=T_{\R\S}\P_\S. \eea
From (\ref{Rdot}) it follows that $\alpha_*=0$ and therefore from (\ref{Tderiv}),
$T_{\S\S}$ and $T_{\R\S}$ have exactly the same $k$ dependence. Taking the derivatives of
(\ref{straightcr}) with respect to wavenumber shows that
\bea n_\C=n_\S\,,  \\ \alpha_\C=\alpha_S\,, \eea
and these consistency relations are true to all orders in slow roll. We can check
explicitly that it holds up to second order in slow roll from Eqs.~(\ref{nC}) and
(\ref{nS}).

The tilts (\ref{nR}) and (\ref{nS}) simplify somewhat to give
\bea  \label{nRstraight} n_\R & \cong & -(6-4\cs)\e+2\ss\esigsig+2\cs\ess +
\(-\frac{10}{3}-4\cs+C\(24-16\cs\)\)\e^2 \\ \nn && +
 \frac23\ss\esigsig^2+\frac23\cs\ess^2
+\(-2+\frac{14}{3}\cs+C\(-16+12\cs\)\)\e\esigsig -\frac23\cs(-1+6C)\e\ess\\
\nn && +\frac23(1+3C)\ss\xisigsigsig+\frac23(1+3C)\cs\xisigss\,,
 \\ \label{nSstraight} n_\S & \cong & -2\e+2\ess +\(-\frac{22}{3}+8C
\)\e^2+\frac23\ess^2+\(\frac83-4C\)\e\esigsig+\(\frac23-4C\)\e\ess
+\frac23(1+3C)\cs\xisigss\,, \eea
and the running is given by
\bea \label{alphaRstraight}
\alpha_\R&\cong&8(-3+4\cs-2\cf)\e^2+4\ss\cs\esigsig^2+4\ss\cs\ess^2+4(4-7\cs+4\cf)\e\esigsig \\
\nn && +4 \cs(5-4\cs)\e\ess-8\ss\cs\esigsig\ess-2\ss\xisigsigsig-2\cs\xisigss\,,
\\ \label{alphaSstraight}
\alpha_\S&\cong&-8\e^2+4\e\esigsig+4\e\ess-2\xisigss\,. \eea

\subsection{Inflaton scenario}

If the background trajectory is straight during inflation then this direction in field
space, $\sigma$, can be identified as the inflaton field. Other fields orthogonal to the
inflaton are time-independent along the background trajectory, but quantum fluctuations
in light fields ($\ess<1$) do generate isocurvature perturbations during inflation. If we
further assume that these isocurvature perturbations remain decoupled from the inflaton
and radiation density during and after inflation then the curvature perturbations today
are purely due to inflaton field perturbations, i.e., $\Trs=0$. We refer to this as the
inflaton scenario. Hence any isocurvature perturbations that survive into the radiation
era are uncorrelated with the curvature perturbations, $\C_{\R\S}=0$ and $\c\!=0$.

The tilts in the inflaton scenario reduce to \bea  \label{nRinflaton} n_\R & \cong &
-6\e+2\esigsig  +
 \(-\frac{10}{3}+24C\)\e^2 +\frac23\esigsig^2
-\(2+16C\)\e\esigsig +\(\frac23+2C\)\xisigsigsig\,,
 \\ \label{nSinflaton} n_\S &\cong & -2\e+2\ess +\(-\frac{22}{3}+8C
\)\e^2+\frac23\ess^2+\(\frac83-4C\)\e\esigsig+\(\frac23-4C\)\e\ess
+\frac23(1+3C)\xisigss\,, \eea
and the running is given by
\bea \label{alphaRinflaton} \alpha_\R&\cong&-24\e^2+16\e\esigsig -2\xisigsigsig\,,
 \\ \label{alphaSinflaton}
\alpha_\S&\cong&-8\e^2+4\e\esigsig+4\e\ess-2\xisigss\,. \eea

The results for the curvature perturbation spectrum in the inflaton case is identical to
the standard result for single-field inflation \cite{stewartlyth}. Since the curvature
perturbations are frozen in from Hubble exit in this case, and the gravitational waves
also are frozen in, the tensor-scalar ratio is also unchanged from the single field case
\bea r\cong 16\e\left[1-\(\frac43+4C\)\e+\(\frac23+2C\)\esigsig\right]\,, \eea and the
standard single field consistency relations apply, see \cite{cortes}
\bea r&\cong & -8n_T \left[1-\frac12n_T+n_\R\right]\,. \eea
Note that by differentiating this expression we can obtain an infinite hierarchy of
consistency relations at higher order in the slow-roll parameters \cite{cortes}, the
first of which is \cite{kosowsky, Lidsey:1995np} \bea \alpha_T & \cong &
n_T\(n_T-n_\R\)\,. \eea

\subsection{Curvaton scenario}\label{curvatonscenario}

In the curvaton scenario \cite{curvaton} the primordial curvature perturbation during the
radiation dominated era is generated from isocurvature field fluctuations in a curvaton
field during inflation. These curvaton perturbations lead to an inhomogeneous radiation
density after inflation if the curvaton is a weakly coupled, massive scalar field whose
energy density is non-negligible when it finally decays into radiation some time after
inflation.

If the effect of curvaton perturbations is assumed to dominate over the contribution from
the inflaton, our chosen normalisation of the dimensionless isocurvature perturbations
during inflation (\ref{defSstar}), implies that $|T_{\R\S}|\gg1$. Thus we have
$\cos\Delta=\pm1$ in Eq.~(\ref{correlationangle}).
The scalar tilt (\ref{nR}) and running (\ref{alphaR}) are therefore
\bea
 n_\R &\cong & -2\e+2\ess +\(-\frac{22}{3}+8C \)\e^2+\frac23\ess^2 +
 \(\frac83-4C\)\e\esigsig+\(\frac23-4C\)\e\ess
+\frac23(1+3C)\xisigss\,, \\
 \alpha_\R & \cong&-8\e^2+4\e\esigsig+4\e\ess-2\xisigss\,.
\eea

If the inflaton perturbation, i.e.~the scalar perturbation at Hubble exit, is negligible,
then from Eq.~(\ref{rstar}) so is the gravitational wave background. We see that for
$\sin\Delta=0$ Eq.~(\ref{rfinal}) implies $r=0$.

Isocurvature perturbations offer an alternative consistency relation in the curvaton
scenario. Not only the primordial curvature perturbation but also any residual
isocurvature perturbation \cite{LUW} is due to the curvaton field fluctuations during
inflation. Thus the primordial curvature and isocurvature perturbations are 100\%
correlated (or anti-correlated), and we must have
\bea n_\R = n_C = n_\S \,,\\
\alpha_\R = \alpha_C = \alpha_\S \,. \eea

\subsection{Curvature perturbations from broken symmetries}

In \cite{kolb} it was proposed that the curvature perturbation seen today may be due to
the isocurvature perturbations during inflation in a nearly symmetric potential. The
isocurvature perturbations are then converted into curvature perturbations during instant
preheating at the end of inflation \cite{felder}. The key parameter of the preheating is
the minimum distance to the minimum of the potential reached along the inflaton
trajectory and perturbations of this minimum distance are due to the isocurvature
perturbations.

The nearly symmetric 2-field potential considered in \cite{kolb} is
 \be
 V(\phi_1,\phi_2)=\frac{m^2}{2}\left(\phi_1^2+\frac{\phi_2^2}{1+x}\right)\,,
  \ee
where $x$ is the symmetry breaking parameter which is assumed to satisfy $0\leq x\ll 1$.
We will assume the symmetry is only weakly broken and so $x$ is of the same order as the
slow roll parameters or less. Hence we will neglect terms of order $x\epsilon^2$ and
$x^2\epsilon$.

The values of the slow-roll parameters depend on the angle of the background trajectory
which is a free parameter of the theory. The distance from the minimum of the potential
when observable scales today were leaving the horizon during inflation (i.e.~about 60
efoldings before the end of inflation) is fixed at $\sigma=\sqrt{\phi_1^2+\phi_2^2}\simeq
3M_{Pl}$. The $\phi_1$ and $\phi_2$ field values can be related to the differential
rotation angle (\ref{thetarotation}), defined by (\ref{tantheta}), by
 \bea
  \phi_1\simeq \sigma\cos\theta(1-x\sin^2\theta)\,, \qquad
 \phi_2\simeq \sigma\sin\theta(1+x\cos^2\theta)\,.
  \eea
These can be derived by applying slow-roll approximations to the equations of motion
 \bea
 \ddot{\phi}_1+3H\dot{\phi}_1+m^2\phi_1=0\,, \qquad
 \ddot{\phi}_2+3H\dot{\phi}_2+\frac{m^2}{1+x}\phi_2=0\,.
 \eea
Note that in the case of a symmetric potential, ($x=0$), the background trajectory is
straight with a polar angle coinciding with $\theta$ \cite{byrnes}.

The potential can be written as
 \bea
V \simeq \frac{m^2}{2}\sigma^2(1-x\sin^2\theta)\,. \eea
The slow-roll parameters are therefore
 \bea
\epsilon = \frac{1}{4\pi G} \frac{1}{\sigma^2}\,, \qquad \esigsig=\epsilon\,, \qquad
\esigs\simeq-x\epsilon\sin\theta\cos\theta\,, \qquad \ess\simeq\epsilon(1-x\cos
2\theta)\,,\qquad \xisigsigsig=\xisigsigs=\xisigss=0\,. \eea

If we suppose that the isocurvature perturbation during inflation dominates the
primordial curvature perturbation we have $\c\!=1$ and the spectral tilt is
 \bea n_\R\cong-2x\epsilon\cos 2\theta-\frac{10}{3}\epsilon^2\,.
 \eea
Note that it is quite possible for the second-order in slow-roll contribution to dominate
over the first-order result.
In the limiting case of a symmetric potential, $V=m^2\sigma^2/2$, there is no
contribution at first order and the tilt is $n_\R=-10\epsilon^2/3$ as calculated in
\cite{byrnes} using another method.

\section{Conclusions}\label{conclusions}

We have calculated the primordial curvature and isocurvature perturbations for an
arbitrary model of two field inflation including first-order corrections in the slow-roll
expansion. We have calculated the power spectra and the cross-correlation, including the
first order corrections [Eqs.\,(\ref{PRfinal}), (\ref{CRSfinal}) and~(\ref{PSfinal})]
which allows us to calculate the spectral tilts to second order in slow roll
[Eqs.\,(\ref{nR}), (\ref{nC}) and~(\ref{nS})]. We find as expected that the scale
dependence of the curvature and isocurvature power spectra are small, being first order
in slow roll. However the scale-dependence of the cross-correlation is not necessarily
small. It becomes large for $\esigs>\c$, where the slow-roll parameter $\esigs$, defined
in appendix \ref{slowrolldefinition}, determines the curvature of the trajectory in field
space. Similarly we find that the running of the curvature and isocurvature tilt is very
small, second order in slow roll, and thus the scale-dependence is well-fit by a
power-law. However the scale dependence of the cross-correlation is not well-described by
a power-law if $\esigs>\c$.

We use the approach of Gordon et al.~\cite{gordon} to evolve the instantaneous adiabatic
and entropy field perturbations on super-Hubble scales. This approach was generalized to
an arbitrary number of fields in curved field space by Nibbelink and van Tent
\cite{nibbelink,vanTent}, who also calculated the leading slow-roll corrections to the
primordial curvature and isocurvature power spectra, making some assumptions about the
evolution of the universe after inflation. The isocurvature perturbations can be evolved
independently of the curvature perturbations on large scales, but the curvature
perturbation can be altered by non-adiabatic perturbations. The primordial power spectra
are thus dependent upon two transfer functions, $T_{\R\S}$ and $T_{\S\S}$, whose
magnitude we leave arbitrary, but whose scale-dependence can be determined in terms of
the slow-roll parameters at Hubble-exit during inflation \cite{wands}.

However on small scales (at early times during inflation) the coupling between the
instantaneous adiabatic and entropy modes can become large leaving the initial vacuum
state ambiguous in terms of these variables. Instead we defined an orthonormal basis in
field space in which the interaction between fields is negligible at early times and
minimises the interaction for a given mode at Hubble-exit. In this basis the effective
mass matrix is diagonalised and the field perturbations are uncorrelated at Hubble exit
at first order in slow roll. But we find that the instantaneous adiabatic and entropy
field perturbations of Gordon et al.~\cite{gordon} {\em are} correlated at Hubble exit at
first order in slow roll, $\c_*\propto\esigs$, unless the trajectory in field space is a
straight line. Correlations between perturbations during 2-field inflation was also
studied in \cite{bmr}.

In \cite{bmrconsistency,wands} it was shown that the well known single-field consistency
relation $r\simeq -8n_T$ can be generalized to the case of two-field inflation to give
$r\simeq -8n_T\ss$. Including the next-order correction terms the two-field consistency
relation becomes
\bea \label{consistency1} r&\cong& -8n_T \ss
\left[1-\frac12n_T+\frac{1}{\ss}n_\R-2\frac{\cs}{\ss}n_\C+\frac{\cs}{\ss}n_\S\right]\,.
\eea

At leading order in slow roll with two fields, there are four slow-roll parameters
($\epsilon$, $\esigsig$, $\esigs$ and $\ess$), and two transfer functions ($T_{\R\S}$ and
$T_{\S\S}$) as well as the energy scale (or Hubble parameter) at Hubble exit during
inflation. However there are, in principle, eight observables, corresponding to the
primordial curvature and isocurvature power spectra, their cross-correlation spectrum,
the tensor spectrum and the tilts of all four spectra. Thus there is one consistency
relation \cite{wands}.  At second-order we have three more slow-roll parameters
($\xisigsigsig$, $\xisigsigs$ and $\xisigss$). Although $\xisss$ is in general non-zero
it does not appear in the calculation of the power spectra, because whenever we take a
time derivative (or derivative with respect to scale) we are differentiating along the
background trajectory, and therefore with respect to the adiabatic field, $\sigma$ and
not $s$.  But we also have four more observables at this order, corresponding to the
running of the four spectra. And thus we have one more consistency relation.  Of course,
there is no guarantee that either the primordial isocurvature perturbations or
gravitational waves will be large enough to be detected.

We can differentiate Eq.~(\ref{consistency1}) with respect to wavenumber, to find a
higher-order consistency relation in terms of the running of the tensor spectrum, see
\cite{vanTent},
 \bea\label{consistency2}
\alpha_T\ss &\cong & n_T\left[n_T-n_\R+\cs(2 n_\C- n_\S-n_T) \right]\,.
 \eea
We could include the third-order correction to this consistency relation by including the
second-order correction term in (\ref{consistency1}), but the result would be extremely
long. Since (\ref{consistency1}) holds on all scales we could repeatedly differentiate it
to calculate an infinite hierarchy of consistency relations, as done in the single field
case \cite{cortes}.

In the case of totally correlated curvature and isocurvature perturbations (i.e.~$\c=0$)
the first consistency relation (\ref{consistency1}) is trivially satisfied since both
$r=0$ and $\s=0$. This corresponds for example to the curvaton scenario discussed in
section \ref{curvatonscenario}. However in this case there is an alternative set of
consistency relations, $n_\S=n_\C=n_\R$ at first order and
$\alpha_\S=\alpha_\C=\alpha_\R$ at second order and so on to give an alternative infinite
hierarchy of relations.

Finally note that observational constraints on the tensor-scalar ratio in the single
field case directly constrain $\e$ to be small, and then the near scale invariance of the
scalar tilt also constrains $\eta$ to be small. Hence higher-order slow-roll corrections
in single field inflation are constrained to be very small. However in two-field
inflation the upper bound on the tensor-scalar ratio does not provide a direct constrain
on $\e$, and the near scale invariance of the adiabatic power spectrum only constrains a
certain combination of the slow-roll parameters to be small. Hence the higher-order
slow-roll corrections in two-field inflation can be significant.

\acknowledgments

The authors are grateful to Jussi V\"{a}liviita for comments. The authors are grateful to
Bartjan van Tent for drawing our attention to related results in \cite{vanTent} and
pointing out a correction to Eq.~(\ref{consistency2}) in v1 of this paper. The authors
are grateful to Ki-Young Choi for pointing out missing terms in the slow roll
derivatives, Eqs.~(\ref{etasigsigdot}--\ref{deltasigHdot}) in v2 of this paper. This
leads to correction at second order in slow roll to several equations, most importantly
for the tilts (\ref{nR}--\ref{nS},\ref{nado},\ref{nadt}), and the runnings
(\ref{alphaR}--\ref{alphaS}). CB acknowledges financial support from the EPSRC.

\appendix

\section{The slow-roll parameters}
\label{slowrolldefinition}\label{relating}

Different papers use different definitions of the slow-roll parameters, some of which are
equivalent at first order, but none are
equivalent at second order. The only ones which we shall use are \\
\indent{First order Hubble}
\begin{eqnarray}
&&\eh=-\frac{\dot{H}}{H^2}=\frac{\m}{2}\frac{\dot{\sigma}^2}{H^2}=\eh_{11}+\eh_{22}\,, \\
&&\eh_{IJ}=\frac{\m}{2}\frac{\dot{\phi}_I\dot{\phi}_J}{H^2}\,,
\\ &&\delta_I^H=-\frac{\ddot{\phi_I}}{H\dot{\phi_I}}\,,
\\ &&\delta_\sigma^H=-\frac{\ddot{\sigma}}{H\dot{\sigma}}\,.
\end{eqnarray}
\indent{First order potential}
\begin{eqnarray}
&& \epsilon=\frac{1}{16\pi G}\left(\frac{V_\sigma}{V}\right)^2\,, \\
&& \eta_{\sigma s}=\frac{1}{\m}\frac{V_{\sigma s}}{V}\,, \qquad
\eta_{\sigma\sigma}=\frac{1}{\m}\frac{V_{\sigma\sigma}}{V}\,, \qquad
\eta_{ss}=\frac{1}{\m}\frac{V_{ss}}{V}\,.
\end{eqnarray}
\indent{Second order potential}
\begin{eqnarray}
&&\xisigsigsig=\frac{1}{(\m)^2}\frac{V_{\sigma\sigma\sigma}V_{\sigma}}{V^2}\,, \qquad
\xisigsigs=\frac{1}{(\m)^2}\frac{V_{\sigma\sigma s}V_{\sigma}}{V^2}\,,
\\ &&\xisigss=\frac{1}{(\m)^2}\frac{V_{\sigma
ss}V_{\sigma}}{V^2}\,, \qquad\,\,\,\,
\xisss=\frac{1}{(\m)^2}\frac{V_{sss}V_{\sigma}}{V^2}\,.
\end{eqnarray}
Note that the superscript 2 refers to the quantity being a second order slow-roll
parameter.

Slow-roll parameters defined in terms of the Hubble parameter can be related to the
potential slow-roll parameters by

\begin{eqnarray} && \eh\cong
\epsilon-\frac43\epsilon^2+\frac23\epsilon\eta_{\sigma\sigma}\,, \\
&& \delta_{\sigma}^H \cong -\epsilon+\eta_{\sigma\sigma}+\frac83\epsilon^2
-\frac83\epsilon\eta_{\sigma\sigma}+\frac13\eta_{\sigma\sigma}^2 +\frac13\esigs^2 +\frac13\xisigsigsig\,, \\
&& \delta_I^H \cong -\eh+A_I+\frac13A_I(A_I-\e)-\frac13\e
A_I-\frac13\frac1H\dot{\delta}_I^H\,,
\end{eqnarray}
where
\bea &&A_1=\eta_{11}+\frac{\sin\theta}{\cos\theta}\eta_{12}\,,
\qquad A_2=\eta_{22}+\frac{\cos\theta}{\sin\theta}\eta_{12}\,, \\
&& A_1+A_2=\eta_{11}+\frac{\eta_{12}}{\sin\theta\cos\theta}+\eta_{22}=2\esigsig+
\esigs\frac{\cos^2\theta-\sin^2\theta}{\sin\theta\cos\theta}\,. \eea

\section{Derivatives}
\label{derivatives}

Note that all slow-roll parameters can only be taken to be constant at the order at which
they are defined, i.e., the time derivative of a first-order slow roll parameter is
second order.
\begin{eqnarray}
\label{epsilondot} && \frac1H\dot{\epsilon}\cong2\epsilon(2\epsilon-\esigsig)\,,
\\ && \label{etasigsigdot}
\frac1H\dotesigsig\cong2\epsilon\esigsig-2\esigs^2-\xisigsigsig\,,
\\ && \label{etasigsdot} \frac1H\dotesigs \cong 2\e\esigs+\esigs(\esigsig-\ess)-\xisigsigs\,,
\\ && \label{etassdot} \frac1H\dotess\cong2\epsilon\ess+2\esigs^2-\xisigss\,, \\&&
\label{deltasigHdot} \frac1H\dot{\delta}_\sigma^H \cong
-4\epsilon^2-2\esigs^2+4\epsilon\esigsig-\xisigsigsig.
\end{eqnarray}
All of the relations above will also be true to the same order if $\frac1H\frac{d}{dt}$
is replaced by $\frac{d}{d\ln k}$ at $k=aH$.

The derivative of $\eh$ is also required to third order for calculating the spectral tilt
of $\Pzero$ (\ref{Pzero}) to second order in slow roll, and for calculating $\beta_*$
(\ref{beta}) to second order in slow roll,
\begin{equation}\label{dlogeh} \frac{d\ln \eh}{d\ln
k}\cong 2 \(2\epsilon-\eta_{\sigma\sigma}+\frac73\epsilon\eta_{\sigma\sigma}
-2\epsilon^2-\frac13\eta_{\sigma\sigma}^2-\frac13\esigs^2-\frac13\xisigsigsig\).
\end{equation}

\section{Calculating the tilt}\label{tiltcalculation}
In order to calculate the spectral indices of the power spectra
(\ref{PRfinal},\ref{CRSfinal},\ref{PSfinal}) we need to calculate the scale dependence of
$\Pzero$, the transfer functions and $a_1,a_2,a_3$, (\ref{aI}), all up to second order in
slow roll.

We define \bea n_{(0)}\equiv \frac{d \log \Pzero}{\dk} \eea where $\Pzero$ is defined by
(\ref{Pzero}) and split the result into first and second order parts
\bea \qquad \label{nstar} \none=-6\e+2\esigsig, \qquad \ntwo
=\frac{14}{3}\e^2+\frac{2}{3}\esigsig^2+\frac23\esigs^2-6\e\esigsig+\frac23\xisigsigsig\,.
\eea

The scale dependence of the transfer functions (\ref{Tderiv}), where the functions
$\alpha_*$ and $\beta_*$ are given by (\ref{alpha},\ref{beta}) respectively and using
$d\log k\simeq H(1-\eh)dt_*$ are
\bea \label{dTss1} \frac{\partial \log T_{\S\S}}{d\log k}\1&=& 2\e-\esigsig+\ess\,, \\
\label{dTss2} \frac{\partial \log \Tss}{\partial\log k}\2&=&
-2\e^2-\frac13\esigsig^2+\frac13\ess^2
+\frac73\e\esigsig+\frac13\e\ess-\frac13\xisigsigsig+\frac13\xisigsigs\,,
\\ \label{dTrs1}
\frac{\partial\log \Trs}{\partial\log k}\1 & = &2\epsilon-\esigsig+2\t \esigs+\ess\,, \\
\label{dTrs2} \frac{\partial\log\Trs}{\partial\log k}\2 &=&
-2\epsilon^2-\frac13\esigsig^2-4C\ts\esigs^2
+\frac13\ess^2+\frac73\epsilon\esigsig-\t\(\frac23+4C\)\epsilon\esigs+
\frac13\epsilon\ess \\ \nn &&+\t\(\frac23+2C\)\esigsig\esigs+\t(\frac23-2C)\esigs\ess
-\frac13\xisigsigsig+\frac23\t\xisigsigs+\frac13\xisigss\,. \eea

Note that the $a_I$, (\ref{aI}) are first order in slow roll, so their derivatives are
second order in slow roll and can be easily calculated from
(\ref{epsilondot},\ref{etasigsigdot},\ref{etasigsdot},\ref{etassdot}).

In order to write the spectral indices in terms of observables we need to be able to
relate the transfer functions and the correlation angle (\ref{correlationangle}). From
(\ref{PRfinal},\ref{CRSfinal},\ref{PSfinal}) it follows that at zeroth order \bea
\label{correlationangle0} && \cos\D\0=\frac{T_{\R\S}}{\sqrt{1+T_{\R\S}^2}}\,, \qquad
\sin\D\0=\frac1{\sqrt{1+T_{\R\S}^2}}\,, \qquad \tan\D\0=\frac1T_{\R\S}\,, \eea and at
first order \bea \label{cosD1}
 && \cos\D\1=\cos\D\0\left[-\frac12a_1 \ss+a_2\frac{\sc}{\c}+\frac12 a_3 \ss\right]\,, \\
\label{sinD1} && \sin\D\1\equiv
\sin\D\0\left[\frac12a_1\cs-a_2\s\c-\frac12a_3\cs\right]\,.
 \eea
Another useful equation for $T_{\R\S}$ is
\bea \label{TRSinverse} \frac{1}{\Trs}=\tan\D\0 \simeq
\t\(1+C\(-2\e+\esigsig-2\esigs\t-\ess\)\)\,. \eea

To find the spectral tilts we take the log derivatives of
(\ref{PRfinal},\ref{CRSfinal},\ref{PSfinal}) and substitute away the dependence on the
transfer functions by replacing them with the correlation angle
(\ref{correlationangle0},\ref{cosD1},\ref{sinD1},\ref{TRSinverse}) to find
\bea n_\R\1&=&\none+2\s\c\dTo\,, \\n_\R\2&=&
\ntwo +2\s\c\dTt-2(\c\s\1+\s\c\1)\dTo  \\ \nn && -2a_1\sc\c\dTo+2a_2\ss(\ss-\cs)\dTo \\
\nn &&+ 2a_3\sc\c\dTo+\ss \dao+2\s\c \dat +\cs \dathree\,,
\\
n_C\1&=& \none+\dlogTsso+\dlogTo\,, \\ n_\C\2&=& \ntwo + \dlogTsst+ \dlogTt+
\t\dat -a_2\t\dlogTo+\dathree\,, \\ n_\S\1 &=& \none+2\dlogTsso\,, \\
n_\S\2&=& \ntwo + 2\dlogTsst+\dathree\,.
 \eea
So substituting (\ref{nstar},\ref{dTss1},\ref{dTss2},\ref{dTrs1},\ref{dTrs2}) in the
above six equations we find the tilts as displayed in (\ref{nR},\ref{nC},\ref{nS}).

\end{document}